\documentclass[12pt]{article}

\pdfoutput=1

\usepackage{graphicx}
\usepackage{epsfig}
\usepackage{amsfonts}
\usepackage{amssymb}
\usepackage{amsmath}
\usepackage{dsfont}
\usepackage{bbm}
\usepackage{multirow}
\usepackage{latexsym}
\usepackage{verbatim}
\usepackage{mcite}
\usepackage{cite}
\usepackage{units}

\def\beq{\begin{equation}}
\def\eeq{\end{equation}}
\def\beqa{\begin{eqnarray}}
\def\eeqa{\end{eqnarray}}

\def\a{{\alpha}}
\def\b{{\beta}}

\def\g{{\gamma}}
\def\bg{{\bar{\gamma}}}

\def\d{{\delta}}
\def\eps{{\epsilon}}

\def\m{{\mu}}
\def\n{{\nu}}
\def\r{{\rho}}
\def\s{{\sigma}}

\def\dm{{\dot{\mu}}}
\def\dn{{\dot{\nu}}}

\def\bfone{\relax{\rm 1\kern-.35em 1}}

\newcommand{\cR}{{\cal R}}
\newcommand{\cS}{{\cal S}}
\newcommand{\cP}{{\cal P}}
\newcommand{\cQ}{{\cal Q}}
\newcommand{\cC}{{\cal C}}
\newcommand{\cM}{{\cal M}}
\newcommand{\cN}{{\cal N}}

\newcommand{\be}{\begin{equation}}
\newcommand{\ee}{\end{equation}}
\newcommand{\ben}{\begin{displaymath}}
\newcommand{\een}{\end{displaymath}}
\newcommand{\bea}{\begin{eqnarray}}
\newcommand{\eea}{\end{eqnarray}}

\newcommand{\bean}{\begin{eqnarray*}}
\newcommand{\eean}{\end{eqnarray*}}

\begin{document}
\pagestyle{plain}


\makeatletter
\@addtoreset{equation}{section}
\makeatother
\renewcommand{\thesection}{\arabic{section}}
\renewcommand{\thefootnote}{\arabic{footnote}}


\setcounter{page}{1}
\setcounter{footnote}{0}


\begin{titlepage}
\begin{flushright}
\small ~~
\end{flushright}

\bigskip

\begin{center}

\vskip 0cm

{\LARGE \bf How to halve maximal supergravity} \\[6mm]

\vskip 0.5cm

{\bf Giuseppe Dibitetto,\, Adolfo Guarino \,and\, Diederik Roest}\\

\vskip 25pt

{\em Centre for Theoretical Physics,\\
University of Groningen, \\
Nijenborgh 4, 9747 AG Groningen, The Netherlands\\
{\small {\tt \{g.dibitetto , j.a.guarino , d.roest\}@rug.nl}}} \\

\vskip 0.8cm

\end{center}

\vskip 1cm

\begin{center}

{\bf ABSTRACT}\\[3ex]

\begin{minipage}{13cm}
\small

We work out the truncation from maximal to half-maximal supergravity in four dimensions. In particular, we determine the explicit constraints on the embedding tensors of both theories. These tensors specify the complete theories, including gauge groups and scalar potentials. Firstly, we find the linear constraint on $\,\cN = 8\,$ theories to allow for a truncation to $\,\cN = 4$. Secondly, we determine the additional $\,\cN = 4\,$ quadratic constraints following from $\,\cN = 8$. Finally, we comment on a brane interpretation as tadpole conditions for the latter.

\end{minipage}

\end{center}

\vfill

\end{titlepage}



\section*{Motivation: a string / supergravity puzzle}
\label{sec:motivation}

Extended supergravity ($\cN \ge 2$) arises as the low energy effective action when dimensionally reducing string theory down to four space-time dimensions while preserving multiple supersymmetries. For instance, maximal supergravity ($\cN = 8$) comes out from \mbox{type II} string compactifications when the internal space is taken to be a six dimensional torus whereas half-maximal supergravity ($\cN = 4$) results as orientifolds thereof. On the string side, the orientifold creates O-planes and eventually also D-branes may be required in order to obtain a consistent compactification. This is normally the case when certain background fluxes are additionally turned on threading the internal geometry and then flux-induced tadpoles for the gauge fields in the R-R sector of the string spectrum have to be cancelled. On the supergravity side, these type II orientifold compactifications including O-planes, D-branes and background fluxes  correspond to half-maximal gauged supergravities, i.e.~deformed supergravity theories in which a certain subgroup of the global symmetry group is realised as a gauge symmetry.

In the last decade a unified framework to describe gauged supergravities has been developed, the so-called embedding tensor formalism \cite{Nicolai:2000sc,deWit:2002vt,deWit:2005ub}. It has been very successful when it comes to study and classify gauged supergravities in a covariant and systematic way, especially in the cases of maximal \cite{deWit:2007mt} and half-maximal \cite{Schon:2006kz} supergravities in four dimensions. Furthermore, the connection between this formalism and string theory realisations has been investigated both for $\cN = 4$ in e.g.~\cite{Aldazabal:2008zza,Dall'Agata:2009gv,Dibitetto:2010rg} and for $\cN = 8$ in e.g.~\cite{deWit:2003hq,Aldazabal:2010ef, Aldazabal:2011yz}. However,
the link between these two cases has received far less attention and indeed some discrepancies seem to appear when matching up different results in the literature.

An example of this already occurs when studying simple type IIA toroidal orientifolds including O6-planes and D6-branes (parallel to the O6-planes) together with gauge and geometric fluxes \cite{Dall'Agata:2009gv}. From the string viewpoint, these compactifications generically give rise to half-maximal supergravities since the O6/D6 sources already break half of the supersymmetries. Even so, an embedding into a maximal supergravity theory could be possible provided that the flux-induced tadpole for the R-R gauge field $C_{7}$ that couples to the O6/D6 sources vanishes. Surprisingly enough, this turns out to happen for the supersymmetric AdS$_{4}$ solutions of these type IIA  orientifold models \cite{Dall'Agata:2009gv} and it has been further extended to the entire set of AdS$_{4}$ solutions in ref.~\cite{Dibitetto:2011gm}. In consequence, one would expect them to actually be solutions of a maximal supergravity theory.

Switching over to the description of the above type IIA orientifolds as half-maximal gauged supergravities, they can be completely classified according to the embedding tensor parameter 
\beq
\begin{array}{ccc}
f_{\a MNP} &\in& \textrm{SL}(2) \times \textrm{SO}(6,6) \ ,
\end{array}
\eeq
where $\,f_{\a MNP}=f_{\a [MNP]}\,$. The indices $\,\a=\pm\,$ and $\,M=1,\dots,12\,$ respectively denote $\,\textrm{SL}(2)\,$ and $\,\textrm{SO}(6,6)\,$ vector indices of the global symmetry group $\,G=\textrm{SL}(2) \times \textrm{SO}(6,6)$. Besides the consistency conditions imposed by half-maximal supergravity, the $f$-tensor has been conjectured to also satisfy
\beq
\label{aldaz_extra}
f_{\a MNP} \, {f_{\b}}^{MNP} = 0 \hspace{10mm} \textrm{ and } \hspace{10mm} f_{\a [MNP} \, {f^{\a}}_{QRS]} = 0
\eeq
for these type IIA orientifolds to admit an embedding into maximal supergravity \cite{Aldazabal:2011yz}. In other words, they would correspond to the consistency conditions imposed by maximal supergravity. A consequence of the second condition in \eqref{aldaz_extra} is the vanishing of total D3-brane charge, which indeed is a necessary condition for maximal supersymmetry.

At this point a string / supergravity puzzle arises since the second condition in (\ref{aldaz_extra}) is \textit{not} satisfied by the AdS$_{4}$ type IIA solutions found in refs~\cite{Dall'Agata:2009gv,Dibitetto:2011gm}, which we expected to be embeddable in a maximal supergravity theory. Even more, the existence of these AdS$_{4}$ solutions is directly related to the non-vanishing of a term in the scalar potential of the form
\beq
\begin{array}{ccc}
V &\supset& f_{\a MNP} \, {f^{\a}}_{QRS} \,\, M^{MNPQRS} \ ,
\end{array}
\eeq
where $\,M^{MNPQRS}=M^{[MNPQRS]}\,$ accounts for the $\textrm{SO}(6,6)$ scalars of the theory. In the type IIA orientifolds we are discussing, this term is sourced by the geometric flux and moreover is needed in order to fix the $\,\textrm{SO}(6,6)\,$ scalars of the theory. This term is present in half-maximal supergravity but, according to the second condition in (\ref{aldaz_extra}), it will no longer be present in maximal supergravity, hence ruling out the AdS$_{4}$ solutions. How can this puzzle be resolved?

Motivated by this puzzle, we have worked out the embedding of a half-maximal supergravity theory into a maximal one in four dimensions. One of the outcomes of the computation is that the second condition in (\ref{aldaz_extra}) -- which represents a six-form of $\,\textrm{SO}(6,6)\,$ -- has to be relaxed to
\beq
\label{aldaz_relax}
\left. f_{\a [MNP} \, {f^{\a}}_{QRS]} \,\, \right|_{\textrm{SD}}= 0 \ ,
\eeq
where the label SD stands for the self-dual part. In contrast, the anti-self-dual (ASD) part still contributes to the scalar potential, 
\beq
\begin{array}{ccc}
V &\supset&  \left. f_{\a [MNP} \, {f^{\a}}_{QRS]} \,\, \right|_{\textrm{ASD}} \,\, M^{MNPQRS} \ ,
\end{array}
\eeq
exactly as it has to in order to recover the AdS$_{4}$ solutions of the type IIA orientifolds. This resolves the apparent contradiction.

\begin{figure}[ht]
\begin{center}
\scalebox{0.55}[0.55]{
\includegraphics[keepaspectratio=true]{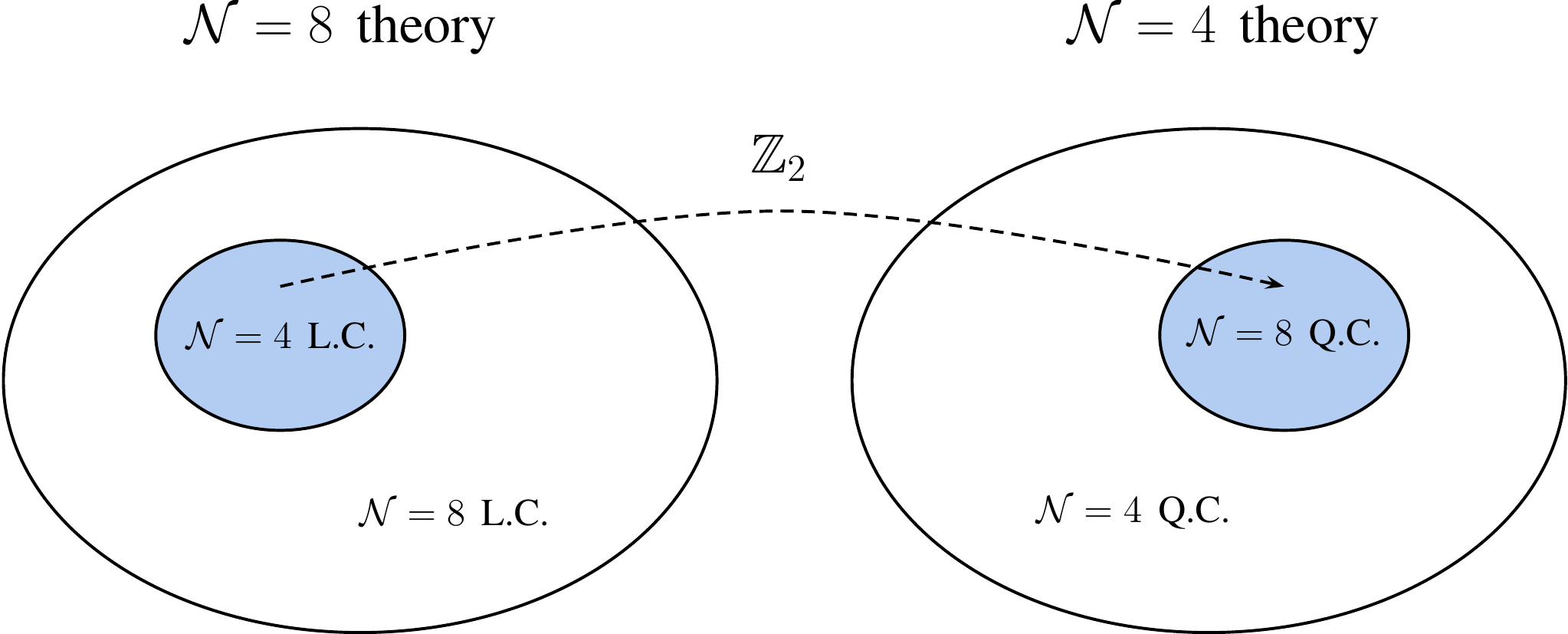} 
}
\end{center}
\caption{The subsets of $\cN = 8$ and $\cN = 4$ theories that can be related via a truncation are indicated. The former subset satisfies a stronger linear constraint (L.C.), while the latter is subject to additional quadratic constraints (Q.C.).}
\label{fig:mapping}
\end{figure}

Irrespective of the puzzle explained above, it is of clear interest to work out the explicit relation between all possible maximal and half-maximal supergravities. This will allow one to determine which $\cN = 8$ gaugings allow to be truncated to $\cN = 4$ and what the explicit relations between the embedding tensors are. Similarly, we will investigate which $\cN = 4$ gaugings follow from a truncation of $\cN = 8$, and hence which consistency conditions are necessary for an uplift to maximal supergravity.  The former involves a number of linear constraints on the $\cN = 8$ embedding tensor, while the latter involves two quadratic constraints on the $\cN = 4$ embedding tensor. See figure~\ref{fig:mapping} for an illustration of these relations. A number of examples of such truncations were discussed in e.g. refs~\cite{Hull:1984wa, Roest:2009tt}, highlighting the fact that electric gaugings of $\cN = 8$ give rise to dyonic gaugings of $\cN = 4$ (in the usual duality frames). With the present more general discussion, we hope to fill in a gap in the literature of extended supergravity and contribute to a better understanding of the link between gauged supergravities and flux compactifications of string theory.

The organisation of this note is as follows. We first discuss the truncation from maximal to half-maximal supergravity stressing the essential role played by the fermionic representations in the decomposition of the $\,\textrm{E}_{7(7)}\,$ tensors of maximal supergravity under the $\,\textrm{SL}(2) \times \textrm{SO}(6,6)\,$ global symmetry of half-maximal supergravity. As a consequence the linear constraint on $\cN = 8$ and the quadratic constraint on $\cN = 4$ will arise. Subsequently, we present the truncation of the scalar sector and demonstrate that this relates the scalar potentials of both theories. We finally offer our closing remarks. The appendix collects some notation and conventions about Majorana-Weyl spinors of $\,\textrm{SO}(6,6)$.

\section*{Truncating maximal supergravity}
\label{sec:N=4_truncation}

In this section we work out the truncation from maximal ($\cN=8$) to half-maximal ($\cN=4$) supergravity in four space-time dimensions. Our departure point is the maximal gauged supergravity framework developed in ref.~\cite{deWit:2007mt}. According to it, the theory can only be deformed by applying \textit{gaugings}, namely, promoting some subgroup $\,G_{0}\,$ of the global symmetry group\footnote{The present results can straightforwardly be extended to include the trombone gauging of ref.~\cite{LeDiffon:2008sh}.} $\,G=\textrm{E}_{7(7)}$ to a local group. The set of possible \textit{gaugings} of maximal supergravity is completely characterised by the embedding tensor parameter $\,{\Theta_{\cM}}^{I}\,$, where $\,\cM=1,\dots,56\,$ and $\,I=1,\dots,133\,$ refer to the fundamental and the adjoint representations of $\,\textrm{E}_{7(7)}\,$ respectively. The embedding tensor determines the way in which the generators $\,X_{\cM}\,$ of the gauge group  $\,G_{0}\,$ are embedded into the $\,\textrm{E}_{7(7)}\,$ generators $\,t_{I}\,$. More specifically,
\beq
\label{X_embedxgener}
X_{\cM} = {\Theta_{\cM}}^{I} \, t_{I} \ ,
\eeq
where (at most) $28$ out of the $56$ vectors $\,X_{\cM}\,$ are linearly independent and enter the \textit{gauging} of the maximal supergravity \cite{deWit:2007mt}. 
The gauge algebra spanned by the vectors $\,X_{\cM}\,$ is written as
\beq
\label{gauge_algebra}
\begin{array}{ccc}
\left[ X_{\cM} , X_{\cN}  \right] = -{X_{\cM\cN}}^{\cP} \, X_{\cP} \ ,
\end{array}
\eeq
where the structure constants (charges) induced by the \textit{gauging} are built using the $\,t_{I}\,$ generators in the fundamental representation, $\,{X_{\cM\cN}}^{\cP} = {\Theta_{\cM}}^{I} \, {{\left[\,t_{I}\,\right]}_{\cN}}^{\cP}\,$. As explained in ref.~\cite{deWit:2007mt}, the $\,\textrm{Sp}(56,\mathbb{R})\,$ invariant skew-symmetric matrix $\,\Omega_{\cM \cN}\,$ can be used  in order to raise and lower $\,\textrm{E}_{7(7)} \subset \textrm{Sp}(56,\mathbb{R})\,$ fundamental indices. From now on, we will adopt the NorthWest-SouthEast (NW-SE) conventions \cite{VanProeyen:1999ni}, e.g.~$\,X_{\cM\cN\cP}={X_{\cM\cN}}^{\cQ}\,\Omega_{\cQ \cP}\,$. 

In order to truncate from maximal to half-maximal supergravity we are making use of the decomposition (branching) of different $\,\textrm{E}_{7(7)}\,$ representations under the $\,\textrm{SL}(2) \times \textrm{SO}(6,6)\,$ symmetry of half-maximal supergravity. Of special interest are
\beqa
\label{irrep_56}
\textbf{56} & \longrightarrow  & (\textbf{2},\textbf{12})\,+\,(\textbf{1},\textbf{32}) \ , \\[2mm]
\label{irrep_133}
\textbf{133} & \longrightarrow  & (\textbf{1},\textbf{66})\,+\,(\textbf{3},\textbf{1})\,+\,(\textbf{2},\textbf{32'}) \ , \\[2mm]
\label{irrep_912}
\textbf{912} & \longrightarrow  & (\textbf{2},\textbf{12})\,+\,(\textbf{2},\textbf{220})\,+\,(\textbf{1},\textbf{352'}) \,+\,(\textbf{3},\textbf{32}) \ , \\[2mm]
\label{irrep_8645}
\textbf{8645}&\longrightarrow & (\textbf{1},\textbf{66})\,+\,(\textbf{1},\textbf{2079})\,+\,(\textbf{3},\textbf{66})\,+\,(\textbf{3},\textbf{495})\,+\,(\textbf{3},\textbf{1})\,+\,(\textbf{1},\textbf{462'}) \,+ \nonumber \\[1mm]
  & &  + \,\,(\textbf{2},\textbf{32'})\,+\,(\textbf{2},\textbf{352})\,+\,(\textbf{2},\textbf{1728'}) \,+\,(\textbf{4},\textbf{32'}) \ ,
\eeqa
where $\textbf{32}$ and $\textbf{32'}$ respectively denote left- and right-handed Majorana-Weyl (M-W) spinorial representations of $\textrm{SO}(6,6)$ and similarly for the other spinorial irrep's\footnote{Notice that the $\textbf{462'}$ irrep in (\ref{irrep_8645}) denotes the SD six-form, which can be built out of right-handed spinor bilinears. See the appendix for conventions about M-W spinorial irrep's of $\,\textrm{SO}(6,6)\,$.}. The decomposition of the $\textbf{56}$ in (\ref{irrep_56}) amounts to the index splitting $\,\cM=(\alpha,M) \,\oplus\, \m\,$, where $\,\alpha=\pm\,$ is an electric-magnetic $\,\textrm{SL}(2)\,$ index, $\,M=1,\dots,12\,$ refers to an $\,\textrm{SO}(6,6)\,$ vector index and $\,\m=1,\dots,32\,$ denotes a M-W left-handed fermionic index. Analogously, an index $\,\dm=1,\dots,32\,$ will denote a M-W right-handed spinor. To carry out the truncation one has to apply a discrete $\,\mathbb{Z}_{2}$-projection\footnote{In a string theory realisation of maximal supergravity, this $\,\mathbb{Z}_{2}$-projection corresponds to orientifolding the theory.} 
\beq
\begin{array}{cccl}
\mathbb{Z}_{2} \,\,: \hspace{5mm}  &       \cN=8           & \longrightarrow &  \cN=4 \\[2mm]     
                                   &  \textrm{E}_{7(7)}    & \longrightarrow &  \textrm{SL}(2) \times \textrm{SO}(6,6) 
\end{array}
\eeq
under which different $\,\textrm{SL}(2) \times \textrm{SO}(6,6)\,$ indices acquire a parity. In particular, the bosonic indices $\,\alpha\,$ and $\,M\,$ are even whereas the fermionic indices $\,\m\,$ and $\,\dm\,$ become odd. Keeping only states which are parity even will truncate from maximal to half-maximal supergravity \cite{Aldazabal:2010ef}.

As a result, the skew-symmetric $\Omega_{\cM \cN}$ matrix becomes block-diagonal with bosonic and fermionic blocks
\beq
\label{Omega}
\Omega_{\cM \cN} =
\left( 
\begin{array}{c|c}
\Omega_{\a M \b N} & 0 \\[1mm] 
\hline
\\[-4mm]
0 & \Omega_{\m \n} 
\end{array}
\right)
= 
\left( 
\begin{array}{c|c}
\eps_{\a \b} \, \eta_{MN} & 0 \\[1mm] 
\hline
\\[-4mm]
0 & \mathcal{C}_{\m \n} 
\end{array}
\right) \ .
\eeq
It is worth observing that the bosonic part involves the Levi-Civita tensor $\,\epsilon_{\alpha \beta}\,$ (with $\,\epsilon_{+ -}=1$) associated to the $\textrm{SL}(2)$ factor as well as the $\,\textrm{SO}(6,6)\,$ metric $\,\eta_{M N}$, whereas the fermionic part only contains the $\,\textrm{SO}(6,6)\,$ invariant antisymmetric matrix $\,\mathcal{C}_{\m \n}$.

The truncation of the charges $\,X_{\cM \cN \cP}\,$ is more subtle. The reason is that they are restricted to sit in the same representation than the embedding tensor $\,{\Theta_{\cM}}^{I}$, namely the $\,\textbf{56} \times \textbf{133}\,$, and furthermore in the $\,\textbf{912}\,$ irrep of $\,\textbf{56} \times \textbf{133}=\textbf{56} + \textbf{912} + \textbf{6480}\,$ because of supersymmetry \cite{deWit:2007mt}. Noticing that $\,\textbf{133}\,$ sits inside $\,\left(\textbf{56} \times \textbf{56}\right)_{s}\,$, this implies the chain  
\beq
\label{rep_chain}
\begin{array}{cccccc}
\textbf{56} \times \left(\textbf{56} \times \textbf{56}\right)_{s} & \longrightarrow &\textbf{56} \times \textbf{133} & \longrightarrow & \textbf{912}  & .
\end{array}
\eeq
This sequence of restrictions on representations can be rephrased in terms of projectors $\,\mathbb{P}^{(133)}\,$ and $\,\mathbb{P}^{(912)}\,$ and hence translates into a set of linear constraints upon $\,X_{\cM \cN \cP}$. In addition to these representation constraints, a set of quadratic constraints remains to be imposed coming from the consistency of the gauge algebra in (\ref{gauge_algebra}). We will now seperately discuss the linear and quadratic constraints.

\subsection*{The linear constraints}

In the truncation to the half-maximal supergravity of ref.~\cite{Schon:2006kz}, the charges $\,X_{\cM \cN \cP}\,$ must be completely specified in terms of the embedding tensor parameters $\,\xi_{\alpha M} \in (\textbf{2},\textbf{12})\,$ and $\,f_{\alpha M N P} = f_{\alpha [M N P]}\in (\textbf{2},\textbf{220})$ which are purely bosonic  and hence parity even with respect to the $\,\mathbb{Z}_{2}$-projection. In other words, the components of the embedding tensor sitting in the $\,(\textbf{1},\textbf{352'})\,$ and $\,(\textbf{3},\textbf{32})\,$ irrep's in the $\,\textbf{912}\,$ decomposition of (\ref{irrep_912}) are parity odd and are projected out. However, this by no means implies that the charges have to be purely bosonic as well. In fact, components of the sorts $\,X_{\a M \b N \g P}\,$, $\,X_{\a M \m \n}\,$, $\, X_{\m \n \a M}\,$ and $\,X_{\m \a M \n}\,$ turn out to be parity even and hence will survive the $\,\mathbb{Z}_{2}$-projection.

Let us start by making the most general antsatz for the $\,X_{\cM \cN \cP}\,$ charges compatible with symmetry in the last two indices, i.e.~they sit in the $\,\textbf{56} \times \left(\textbf{56} \times \textbf{56}\right)_{s}\,$. Due to the definition in (\ref{X_embedxgener}), the charges depend linearly on the embedding tensor parameters. In terms of the $\,\xi_{\a M}\,$ and $\,f_{\a MNP}\,$ components which are invariant under the $\,\mathbb{Z}_{2}$-projector, and hence survive the truncation, they take the form of
\beq
\label{X_antsatz}
\begin{array}{cclc}
X_{\a M \b N \g P} & = & c_{0} \, \epsilon_{\b \g} \, f_{\a MNP} + d_{0} \, \epsilon_{\b \g} \, \eta_{M [N}\, \, \xi_{ \a P]} +  d'_{0} \, \epsilon_{\a (\b} \, \xi_{\g) M} \, \eta_{NP}& , \\[4mm]
X_{\a M \m \n} & = & c_{1} \, f_{\a MNP} \, \left[ \g^{NP} \right]_{\m \n}  +  d_{1} \, \xi_{\a N} \, \left[ {\g_{M}}^{N} \right]_{\m \n}         & , \\[4mm]
X_{\m \a M \n} = X_{\m \n \a M} & = & c_{2} \, f_{\a MNP} \, \left[ \g^{NP} \right]_{\m \n} + c_{3} \, f_{\a NPQ} \, \left[ {\g_{M}}^{NPQ} \right]_{\m \n} \\[3mm]
& + & d_{2} \, \xi_{\a N} \, \left[ {\g_{M}}^{N} \right]_{\m \n} + d_{3} \, \xi_{\a M} \, \cC_{\m \n}        & ,   
\end{array}
\eeq
where $c$'s and $d$'s are constant coefficients to be fixed in the following by imposing representation constraints.

First of all, the charges must sit in the $\,\textbf{56} \times \textbf{133}\,$. Imposing this amounts to require invariance under the action of a $\,\mathbb{P}^{(133)}\,$ projector 
\beq
\label{X_P133_invariance}
X_{\cM \cN \cP} = {{\mathbb{P}^{(133)}}_{\cN \cP}}^{\cQ \cR} \, X_{\cM \cQ \cR} \ ,
\eeq
that keeps only the $\,\textbf{133}\,$ in the decomposition $\,\left(\textbf{56} \times \textbf{56}\right)_{s} = \textbf{133} + \textbf{1463}\,$. The expression for this projector reads
\beq
\label{P133_proj}
{{\mathbb{P}^{(133)}}_{\cM \cN}}^{\cP \cQ} = K^{IJ} \,\, \left[ t_{I} \right]_{\cM \cN} \, \left[ t_{J} \right]^{\cP \cQ} \ ,
\eeq
where $\,K^{IJ}\,$ denotes the inverse of the Killing-Cartan metric
\beq
\label{KC_metric}
K_{IJ}=\textrm{Tr}(t_{I} \, t_{J})=[t_{I}]_{\cM \cN} \, [t_{J}]_{\cP \cQ} \,\, \Omega^{\cM \cQ} \, \Omega^{\cN \cP} \ ,
\eeq
which, in turn, also depends on the $\,[t_{I}]_{\cM \cN}\,$ symmetric generators of $\,\textrm{E}_{7(7)}\,$ in the fundamental representation. By virtue of the branching (\ref{irrep_133}), the general form of the generators is given by
\beq
\label{E7_gen}
\begin{array}{cclc}
\left[t_{\a M \b N}\right]_{\g P \d Q} & = & \epsilon_{\a \b} \, \epsilon_{\g \d} \, \left[t_{MN}\right]_{PQ} + \eta_{MN}\, \, \eta_{PQ} \, \left[t_{\a \b}\right]_{\g \d} & , \\[3mm]
\left[t_{\a M \b N}\right]_{\m \n} & = &\dfrac{1}{4} \, \epsilon_{\a \b} \, \left[\g_{MN}\right]_{\m \n} & , \\[3mm]
\left[t_{\a \dm}\right]_{\b N \n} = \left[t_{\a \dm}\right]_{\n \b N} & = & \epsilon_{\a \b} \, \left[\bg_{N}\right]_{\dm \n} = - \,\epsilon_{\a \b} \, \left[\g_{N}\right]_{\n \dm}& .
\end{array}
\eeq
The normalisation of these generators is irrelevant for the present purposes as it drops out of the definition of the projector. The ratio between the two representations of the generator $t_{\a M \b N}$, however, is important and can be fixed by requiring the two representations to satisfy the same commutation relation.
Using the above form of the $\,\textrm{E}_{7(7)}\,$ generators, the Killing-Cartan metric (\ref{KC_metric}) comes out with a block diagonal structure 
\beq
\label{K_block}
K_{I J} =
\left( 
\begin{array}{c|c}
K_{\a M \b N , \g P \d Q} & 0 \\[1mm] 
\hline
\\[-4mm]
0 & K_{\a \dm , \b \dn} 
\end{array}
\right) \ ,
\eeq
with non-vanishing components
\beq
\begin{array}{cclc}
K_{\a M \b N , \g P \d Q} & = & -12 \, \eta_{MN} \, \eta_{PQ} \,  \epsilon_{\a (\g} \, \epsilon_{\d) \b} - 2  \, \eps_{\a \b} \, \eps_{\g \d} \ \eta_{M [P} \,\eta_{Q] N} &  , \\[2mm]
K_{\a \dm , \b \dn} & = & - 24 \, \epsilon_{\a \b} \, \cC_{\dm \dn} & .
\end{array}
\eeq
The inverse of the Killing-Cartan metric is then also block-diagonal with non-vanishing components\footnote{In the computation of the inverse we have taken $\,[t_{\a \b}]^{\g \d} = \delta_{\a}^{(\g}\delta_{\b}^{\d)}\,$ and $\,[t_{M N}]^{P Q} = \delta_{MN}^{PQ}\,$. This is consistent with the definitions
\beq
K^{\textrm{SL}(2)}_{\a \b , \g \d} \equiv - \eps_{\a (\g} \, \eps_{\d) \b} \hspace{10mm} \textrm{ and } \hspace{10mm} K^{\textrm{SO}(6,6)}_{M N , P Q} \equiv - \eta_{M [P} \,\eta_{Q] N} \ ,
\eeq
of the $\,\textrm{SL}(2)\,$ and $\,\textrm{SO}(6,6)\,$ Killing-Cartan metrics.
}
\beq
\begin{array}{cclc}
K^{\a M \b N , \g P \d Q} & = & -\dfrac{1}{1728} \, \eta^{MN} \, \eta^{PQ} \,  \epsilon^{\a (\g} \, \epsilon^{\d) \b} -  \dfrac{1}{8} \, \eps^{\a \b} \, \eps^{\g \d} \ \eta^{M [P} \,\eta^{Q] N}&  , \\[3mm]
K^{\a \dm , \b \dn} & = & - \dfrac{1}{24} \, \epsilon^{\a \b} \, \cC^{\dm \dn} & .
\end{array}
\eeq
Plugging the above results into the definition of the $\mathbb{P}^{(133)}$ projector in (\ref{P133_proj}), its components take the simple form of
\beq
\begin{array}{cclc}
{{\mathbb{P}^{(133)}}_{\a M \b N}}^{\g P \d Q} & = &  - \dfrac{1}{2} \, \epsilon_{\a \b} \, \epsilon^{\g \d} \, \delta_{MN}^{PQ} + \dfrac{1}{12} \, \delta_{\a}^{(\g} \, \delta_{\b}^{\d)} \, \eta_{M N}\, \eta^{P Q}& , \\[4mm]
{{\mathbb{P}^{(133)}}_{\a M \b N}}^{\m \n} & = & - \dfrac{1}{8} \, \eps_{\a \b} \, \left[ \g_{MN} \right]^{\m \n} & , \\[4mm]
{{\mathbb{P}^{(133)}}_{\m \n}}^{\r \s} & = & - \dfrac{1}{32} \, \left[ \g_{MN} \right]_{\m \n} \, \, \left[ \g^{MN} \right]^{\r \s} & , \\[4mm]
 {{\mathbb{P}^{(133)}}_{\a M \m}}^{\b N \n} & = & \dfrac{1}{24} \, \delta_{\a}^{\b} \, \left( \, {\left[ {\g_{M}}^{N} \right]_{\m}}^{\n}  +  {\delta_{M}}^{N} \, {\delta_{\m}}^{\n} \right)& .
\end{array}
\eeq
Applying it to the charges and requiring (\ref{X_P133_invariance}), we find that
\beq
c_0 = 4 \, c_1 
\hspace{5mm} , \hspace{5mm} 
d_0 = 4 \, d_1
\hspace{5mm} , \hspace{5mm}
c_2 = -3 \, c_3 
\hspace{5mm} \textrm{ and }  \hspace{5mm}
d_2 = - d_3 \ .
\eeq

Finally we still have to impose that the charges $\,X_{\cM \cN \cP}\,$ live in the \textbf{912}. The expression for the $\,\mathbb{P}^{(912)}\,$ projector acting on the embedding tensor $\,{\Theta_{\cM}}^{I}\,$ can be found in ref.~\cite{deWit:2002vt}. However we are interested in the constraints it imposes on the charges. These happen to be  \cite{ deWit:2007mt}
\beq
\label{912_cond}
X_{(\cM \cN \cP)} = 0 \hspace{10mm} \textrm{ and }  \hspace{10mm} {X_{\cM \cN}}^{\cM} = 0 \ ,
\eeq
and translate into additional relations between the coefficients
\beq
c_1 = -2 \, c_2 \hspace{5mm} \textrm{ , }  \hspace{5mm} d_1 = - 2 \, d_2 \hspace{8mm} \textrm{ and }  \hspace{8mm} d_{0}=d'_{0} \ ,
\eeq
all of them coming from the first condition in (\ref{912_cond}). This determines the charges in (\ref{X_antsatz}) up to a global size of the SO$(6,6)$ and the SL$(2)$ embedding tensor parameters. These two global sizes can be set in order to reproduce the conventions in ref.~\cite{Schon:2006kz}.  

The final outcome is then given by
\beq
\label{X}
\begin{array}{cclc}
X_{\a M \b N \g P} & = & - \, \epsilon_{\b \g} \, f_{\a MNP} - \, \epsilon_{\b \g} \, \eta_{M [N}\, \, \xi_{ \a P]} - \, \epsilon_{\a (\b} \, \xi_{\g) M} \, \eta_{NP}& , \\[4mm]
X_{\a M \m \n} & = & -\dfrac{1}{4}  \, f_{\a MNP} \, \left[ \g^{NP} \right]_{\m \n}  -  \dfrac{1}{4} \, \xi_{\a N} \, \left[ {\g_{M}}^{N} \right]_{\m \n}         & , \\[4mm]
X_{\m \a M \n} = X_{\m \n \a M} & = & \dfrac{1}{8} \, f_{\a MNP} \, \left[ \g^{NP} \right]_{\m \n} - \dfrac{1}{24} \, f_{\a NPQ} \, \left[ {\g_{M}}^{NPQ} \right]_{\m \n} \\[3mm]
& + & \dfrac{1}{8} \, \xi_{\a N} \, \left[ {\g_{M}}^{N} \right]_{\m \n} - \dfrac{1}{8} \, \xi_{\a M} \, \cC_{\m \n}        & . \\[2mm]
\end{array}
\eeq
This specifies the relation between the embedding tensor components of half- and maximal supergravity. Only tensors in the $\bf 912$ of this specific form allow for a truncation to $\cN = 4$. At this point we want to stress that the components $\,\xi_{\a M}\,$ and $\,f_{\a MNP}\,$ in the truncated theory will \textit{necessarily} source fermionic components of the charges -- $X_{\a M \m \n}\,$ and $\,X_{\m \a M \n} = X_{\m \n \a M}\,$ by means of contractions with $\,\gamma$-matrices -- whenever half-maximal is embedded into maximal supergravity. As we will see later on, these fermionic components also contribute to the scalar potential of the theory and hence they have to be included in order to consistently truncate from maximal to half-maximal supergravity.

For future use it will be convenient to also give the $\Theta$-components of the embedding tensor. Using (\ref{X_embedxgener}) and the $\,\textrm{E}_{7(7)}\,$ generators in (\ref{E7_gen}), these are given by
\beq
\label{emb_tens_comp}
\begin{array}{cclc}
{\Theta_{\a M}}^{\b N \g P} & = & - \dfrac{1}{2} \, \epsilon^{\b \g} \, {f_{\a M}}^{NP} \, -  \, \dfrac{1}{2} \, \epsilon^{\b \g}  \, \delta_{M}^{[N} \, \xi_{\a}^{\,\,\,\,P]} \, + \,  \dfrac{1}{12} \,  \delta_{\a}^{(\b} \, \xi^{\g)}_{\,\,\,\,M} \, \eta^{N P} & , \\[4mm]
{\Theta_{\m}}^{\a \dn} & = & \dfrac{1}{24} \, \epsilon^{\a \b} \, f_{\b M N P} \, \left[  \g^{M} \, \bar{\g}^{N} \, \g^{P} \right]_{\mu}^{\,\,\,\,\,\,\dn} \, - \, \dfrac{1}{8} \, \epsilon^{\a \b} \, \xi_{\b M} \left[ \gamma^{M}\right]_{\mu}^{\,\,\,\,\,\,\dn} & .
\end{array}
\eeq

\subsection*{The quadratic constraints}

The closure of the gauge algebra in (\ref{gauge_algebra}) imposes additional constraints on the charges of the maximal supergravity theory. These are quadratic constraints which can be written as
\be
\label{QC8} 
X_{\cM \cN \cP} \,\, X_{\cQ \cR \cS} \,\, \Omega^{\cM \cQ}= {\Theta_{\cM}}^{I} \,  {\Theta_{\cQ}}^{J} \, \Omega^{\cM \cQ} \, [t_{I}]_{\cN \cP} \, [t_{J}]_{\cR \cS} = 0 \ ,
\ee
so they sit in the $\,\left( \textbf{133} \times \textbf{133} \right)_{a} = \textbf{133}\,+\,\textbf{8645}\,$ irrep's of the $\textrm{E}_{7(7)}$ symmetry group of maximal supergravity \cite{ deWit:2007mt}. By virtue of the truncation to half-maximal supergravity, we expect the quadratic constraints solely to furnish the parity even irrep's appearing in the decompositions (\ref{irrep_133}) and (\ref{irrep_8645}). These are\footnote{The last irrep in \eqref{irrep_133} and the ones in the second
line of \eqref{irrep_8645} fit some spinorial representations of SO($6,6$) (to be more precise left-/right-handed spinors, spinor-vector and spinorial 2-form respectively). They happen to be odd under the $\,\mathbb{Z}_{2}$-projection performed in the truncation from maximal to half-maximal theories and hence they are projected out.}$^,$\footnote{The first set of quadratic constraints is empty for vanishing $\,\xi_{\a M}$, while the second is for vanishing $\,f_{\a MNP}$.}
\beqa
\label{irrep_133_Z2}
\textbf{133} & \overset{\mathbb{Z}_2}{\longrightarrow}  & (\textbf{1},\textbf{66})\,+\,(\textbf{3},\textbf{1}) \ , \\[2mm]
\label{irrep_8645_Z2} 
\textbf{8645}& \overset{\mathbb{Z}_2}{\longrightarrow} & (\textbf{1},\textbf{66})\,+\,(\textbf{1},\textbf{2079})\,+\,(\textbf{3},\textbf{66})\,+\,(\textbf{3},\textbf{495})\,+\,(\textbf{3},\textbf{1})\,+\,(\textbf{1},\textbf{462'}) \ . \,\,\,\,\,\,
\eeqa

Plugging the expression for the charges in (\ref{X}) -- or identically the one for the embedding tensor in (\ref{emb_tens_comp}) -- and for the $\,\Omega$-matrix in (\ref{Omega}) inside the quadratic constraints of (\ref{QC8}), one rediscovers the set of quadratic constraints of half-maximal supergravity plus two additional ones which then tell us which is the subset of $\,\cN=4\,$ gaugings which can be viewed as truncations of an $\,\cN=8\,$ theory. The former set of quadratic constraints was derived in ref.~\cite{Schon:2006kz} and consists of the conditions
\beqa
\label{QC41} 
&\hspace{-15mm} i)  & \xi_{\a M}\,\xi_{\b}^{\phantom{a}M}=0 \ , \\[2mm]
\label{QC42}
&\hspace{-15mm} ii) & \xi_{(\a}^{\phantom{a}P}\,f_{\b)PMN}=0 \ , \\[2mm]
\label{QC43}
&\hspace{-15mm} iii)& 3\,f_{\a R[MN}\,f_{\b PQ]}^{\phantom{abcde}R}\,+\,2\,\xi_{(\a [M}\,f_{\b)NPQ]}=0 \ , \\[2mm]
\label{QC44}
&\hspace{-15mm} iv) & \eps^{\a \b}\left(\xi_{\a}^{\phantom{a}P}\,f_{\b PMN}\,+\,\xi_{\a M}\,\xi_{\b N}\right)=0 \ , \\[2mm]
\label{QC45}
&\hspace{-15mm} v) & \eps^{\a \b}\left(f_{\a MNR}\,f_{\b PQ}^{\phantom{abcde}R}\,-\,\xi_{\a}^{\phantom{a}R}\,f_{\b R[M[P}\,\eta_{Q]N]}\,-\,\xi_{\a [M}\,f_{\b N]PQ}\,+\,\xi_{\a [P}\,f_{\b Q]MN}\right)=0 \ , 
\eeqa
which correspond with the following irrep's of the $\,\textrm{SL}(2) \times \textrm{SO}(6,6)\,$ symmetry of half-maximal supergravity,
\beq
\label{irrepQC41}
i) \quad (\textbf{3},\textbf{1})
\hspace{10mm}  \hspace{10mm}
ii) \quad (\textbf{3},\textbf{66})
\hspace{10mm}  \hspace{10mm}
iii) \quad (\textbf{3},\textbf{495})
\eeq
\beq
\label{irrepQC42}
iv) \quad (\textbf{1},\textbf{66})
\hspace{10mm}  \hspace{10mm}
v) \quad (\textbf{1},\textbf{66})\,+\,(\textbf{1},\textbf{2079}) \ .
\eeq
By matching up the irrep's in \eqref{irrepQC41} and \eqref{irrepQC42} with those given in \eqref{irrep_133_Z2} and \eqref{irrep_8645_Z2}, one realises that the two extra conditions to be imposed on $\,f_{\a M N P}\,$ and $\,\xi_{\a M}\,$ in order to describe a truncation from maximal supergravity must correspond with the $\,(\textbf{3},\textbf{1})\,$ and $\,(\textbf{1},\textbf{462'})\,$ irrep's both coming from \eqref{irrep_8645_Z2}. Indeed, they respectively turn out to be
\beq
\label{correct_extra}
f_{\a MNP} \, {f_{\b}}^{MNP} = 0 \hspace{10mm} \textrm{ and } \hspace{10mm} \left. \eps^{\a \b}\,\, f_{\a [MNP} \, f_{\b QRS]} \,\, \right|_{\textrm{SD}} = 0 \ ,
\eeq
where the second condition just picks out the self-dual (SD) part of the whole $\,\textrm{SO}(6,6)\,$ six-form $\,f_{\a [MNP} \, {f^{\a}}_{QRS]}\,$. This feature stems from the fact that the six-form gets contracted with the $\,[\gamma^{MNPQRS}]^{\dm \dn}\,$ matrix -- which turns out to be SD in the case of the $\,\textrm{SO}(6,6)\,$ group, see the appendix for more details -- during the computation of the quadratic constraints.

\subsection*{The scalar sector}
\label{sec:scalar_sector}

We now turn to the truncation of the scalar sector from maximal to half-maximal supergravity by applying the $\,\mathbb{Z}_{2}$-projection introduced in the previous section. Its action on the coset geometry of maximal supergravity,
\beq
\begin{array}{cccl}
\mathbb{Z}_{2} \,\,: \hspace{5mm}  &       \cN=8           & \longrightarrow &  \cN=4 \\[2mm]     
                                   &  \dfrac{\textrm{E}_{7(7)}}{\textrm{SU}(8)}    & \longrightarrow &  \dfrac{\textrm{SL}(2)}{\textrm{SO}(2)} \times \dfrac{\textrm{SO}(6,6)}{\textrm{SO}(6) \times \textrm{SO}(6)} \ ,
\end{array}
\eeq
reduces the number of scalar fields in the truncated theory from $\,70\,$ to $\,38\,$. The parameterisation of the $\,\textrm{E}_{7(7)}/\textrm{SU}(8)\,$ coset is given by a symmetric $\,\mathcal{M}_{\cM \cN}\,$ matrix which, after the truncation to half-maximal supergravity, becomes block-diagonal,
\beq
\label{E7_scalars}
\mathcal{M}_{\cM \cN} = 
\left( 
\begin{array}{c|c}
\cM_{\a M \b N} & 0 \\[1mm] 
\hline
\\[-4mm]
0 & \cM_{\m \n} 
\end{array}
\right)
=
\left( 
\begin{array}{c|c}
M_{\a \b} \, M_{MN} & 0 \\[1mm] 
\hline
\\[-4mm]
0 & \dfrac{1}{6!} \, M_{MNPQRS} \, \left[ \g^{MNPQRS} \right]_{\m \n} 
\end{array}
\right) \ ,
\eeq
with a bosonic $\,\cM_{\a M \b N}\,$ and a fermionic $\,\cM_{\m \n}\,$ block. The former contains the $\,\textrm{SL}(2)\,$ and the $\,\textrm{SO}(6,6)\,$ scalars $\,M_{\a \b}\,$ and $\,M_{MN}\,$ of half-maximal supergravity whereas the latter now involves a contraction with the $\,[\gamma^{MNPQRS}]_{\m \n}\,$ ASD matrix. This time it is contracted with the $\,\textrm{SO}(6,6)\,$ six-form
\beq 
\label{M6}
M_{MNPQRS} \equiv \epsilon_{mnpqrs}\mathcal{V}_{M}^{\phantom{M}m}\mathcal{V}_{N}^{\phantom{M}n}\mathcal{V}_{P}^{\phantom{M}p}\mathcal{V}_{Q}^{\phantom{M}q}\mathcal{V}_{R}^{\phantom{M}r}\mathcal{V}_{S}^{\phantom{M}s} \ ,
\eeq
where $\,\mathcal{V}\,$ denotes an $\,\textrm{SO}(6,6)/\textrm{SO}(6) \times \textrm{SO}(6)\,$ Zw\"{o}lfbein such that $\,M=\mathcal{V}\,\mathcal{V}^{T}\,$ and the index $\,m\,$ only runs over the six time-like directions \cite{Schon:2006kz}.  At the origin of the moduli space one obtains $\,\cM_{\a M \b N}=\delta_{\a \b}\,\delta_{MN} \, \,$ and $\,\cM_{\m \n}= {A^{\r}}_{\n} \, \cC_{\r \m} \equiv B_{\m\n}\,$ where $\,A=\frac{1}{6!} \, \eps_{mnpqrs} \, \g^{mnpqrs} \,$. It is worth noticing here that the symmetric $\,B_{\m\n}\,$ matrix \cite{VanProeyen:1999ni} is the correct one in order for the origin of the moduli space to be invariant under the action of the compact part of the symmetry group in half-maximal supergravity, i.e.~$\,\textrm{SO}(2) \times \textrm{SO}(6) \times \textrm{SO}(6)\,$.

Let's have a look at  the effect of the above decomposition on the kinetic terms of the scalar sector. In particular, the $\textrm{E}_{7(7)}$ kinetic terms give rise to the following, upon insertion of \eqref{E7_scalars},
\begin{align}
e^{-1} \, \mathcal{L}_{kin} \, = \,  \dfrac{1}{96} \, \partial \cM_{\cM \cN} \, \partial \cM^{\cM \cN} \,  =  \, \dfrac{1}{8} \, \partial M_{\a \b} \, \partial M^{\a \b} \, + \, \dfrac{1}{16} \, \partial M_{MN} \, \partial M^{MN} \ .  
\end{align}
It is worth mentioning that the fermionic block contributes to $\,(\partial M_{M N})^2\,$ such that both the $\,\textrm{SL}(2)\,$ and $\,\textrm{SO}(6,6)\,$ kinetic terms are reproduced with the correct normalisation.

A final check of our results is to see whether the scalar potentials, which are completely specified in terms of the embedding tensor for both theories, also naturally relate to each other.  The scalar potential in maximal supergravity takes the form~\cite{ deWit:2007mt}
\beq
\begin{array}{ccc}
\label{V_N=8}
V &=& \dfrac{g^{2}}{672} \, X_{\cM \cN \cP}  \, X_{\cQ \cR \cS} \left( \mathcal{M}^{\cM \cQ} \, \mathcal{M}^{\cN \cR} \, \mathcal{M}^{\cP \cS}   +   7 \, \mathcal{M}^{\cM \cQ} \, \Omega^{\cN \cR} \, \Omega^{\cP \cS}    \right) \ .
\end{array}
\eeq
Plugging the expression for the embedding tensor and the parameterisation of the scalars in the truncated theory, one correctly reproduces the scalar potential of half-maximal supergravity. The general form of the latter is given by \cite{Schon:2006kz} 
\beq
\begin{array}{ccl}
\label{V_N=4}
V & = &  \dfrac{g^{2}}{16} \left\{ \, f_{\alpha MNP} \, f_{\beta QRS} M^{\alpha \beta} \left[ \dfrac{1}{3} \, M^{MQ} \, M^{NR} \, M^{PS} + \left(\dfrac{2}{3} \, \eta^{MQ} - M^{MQ} \right) \eta^{NR} \eta^{PS} \right] \right. \\[4mm]
  & - & \left. \dfrac{4}{9} \, f_{\alpha MNP} \, f_{\beta QRS} \, \epsilon^{\alpha \beta} \, M^{MNPQRS} \,+\, 3 \, \xi_{\alpha M} \, \xi_{\beta N} \, M^{\a \b} \, M^{MN} \right\} \ .
\end{array}
\eeq
After a straightforward but tedious calculation\footnote{It turns out to be crucial for calculational efficiency to first express $\,V\,$ in terms of $\Theta$ rather than $X$ before carrying out the $\gamma$-matrix manipulations.}, we have explicitly checked that indeed the $\,\cN = 8\,$ truncation gives rise to this scalar potential, of course modulo the additional quadratic constraints in \eqref{correct_extra}. For this reason, the term that is independent of $\,\textrm{SO}(6,6)\,$ scalars is in fact not present, and the term that is independent of $\,\textrm{SL}(2)\,$ scalars involves only the anti-self-dual part of the $\textrm{SO}(6,6)$ six-form.

\section*{Concluding remarks}
\label{sec:conclusions}

In the present paper we have addressed the issue of how to consistently truncate from maximal to half-maximal supergravity in four space-time dimensions. In particular, we have shown that spinorial components of the embedding tensor of the maximal theory are crucial when truncating to the half-maximal theory. This is a consequence of the linear constraints encoded in \eqref{X}. The natural interpretation of this constraint is that this corresponds to retaining only the parity even components of the $\,\bf 912\,$ of $\,\textrm{E}_{7(7)}$. Similarly, we have found that $\,\cN = 4\,$ theories can only be embedded in $\,\cN = 8\,$ if they satisfy the two additional quadratic constraints \eqref{correct_extra}.  A natural question is what these additional conditions correspond to.

Using the dictionary between the embedding tensor and fluxes \cite{Dibitetto:2010rg}, it can be seen that the $\,({\textbf 1}, {\textbf{462'}})\,$ includes a component of the form
\begin{align}
\epsilon^{mnpqrs} \, F_{mnp} \, H_{qrs} \ ,
\end{align}
where $\,F_{mnp}\,$ and $\,H_{mnp}\,$ are the R-R and NS-NS three-form internal fluxes of IIB compactifications. Therefore this corresponds to the tadpole condition imposing the vanishing of the total D3-brane charge. Other components of the same irrep impose the absence of net charges of duality related branes. It would be interesting to investigate which other branes fill up the $\,\textrm{SL}(2) \times \textrm{SO}(6,6)\,$ orbit of the D3-brane\footnote{Note that the D3-branes fill out a different $\,\textrm{SL}(2) \times \textrm{SO}(6,6)\,$ orbit in ref.~\cite{Bergshoeff:2010xc}. This stems from different embeddings in $\,\textrm{E}_{7(7)}$, see also ref.~\cite{Aldazabal:2010ef}.}. Similarly, it remains to be seen whether the $\,(\textbf{3},\textbf{1})\,$ additional quadratic constraint also corresponds to a tadpole condition or whether it has some other interpretation. We hope to come back to these issues in the future.

%
%

\section*{Acknowledgments}

We are grateful to E.~Bergshoeff and A.~Borghese for very stimulating discussions. The work of the authors is supported by a VIDI grant from the Netherlands Organisation for Scientific Research (NWO).

%
%

\appendix

\section*{Appendix: Majorana-Weyl spinors of $\textrm{SO}(6,6)$}
\label{App:Maj-Wey_spin}

Majorana spinors in $6+6$ dimensions have $64$ real independent
components. This implies that there exists a purely real
representation for the Dirac matrices $\{\Gamma^M\}_{M=1,\cdots,12}$
such that they satisfy
\bea\label{Clifford}
\left\{\Gamma^M,\Gamma^N\right\}=2\,\eta^{MN}\,\mathds{1}_{64}\,,
\eea
where $\eta^{MN}=\textrm{diag}(-1,\cdots,-1,+1,\cdots,+1)$\,.

We adopt a set of conventions in which spinors are naturally objects
of the form $\chi^{a}$ and hence Dirac matrices carry indices
$\left[\Gamma^M\right]^{a}_{\phantom{a}b}\,$. However, in order to
discuss symmetry and antisymmetry properties of gamma matrices, we
introduce two antisymmetric objects, $\cC_{a b}$ and $\cC^{a b}$,
which turn out to represent the components of the charge conjugation
matrix and its inverse transpose respectively. We will use these
objects in order to raise and lower spinorial indices according to
the so-called NorthWest-SouthEast (NW-SE)
conventions \cite{VanProeyen:1999ni}. This translates into the
following prescriptions
\bea \chi^{a}=\cC^{a b}\,\chi_b\hspace{8mm},\hspace{8mm}
\chi_a=\chi^b\,\cC_{b a}\,. \eea
As explained in ref.~\cite{VanProeyen:1999ni}, the consistency of the two
rules presented above implies
\bea \cC^{a b}\,\cC_{c
b}=\delta^{a}_{\phantom{a}c}\hspace{8mm}\textrm{and}\hspace{8mm}\cC_{b
a}\,\cC^{b c}=\delta_{a}^{\phantom{a}c}\,. \eea
The charge conjugation matrix mentioned above relates Dirac matrices
to their transpose in the following way
\bea\label{GammaT} \left(\Gamma^M\right)^T
=-\,\cC\,\Gamma^M\,\cC^{-1}\,.\eea

However, the $\textbf{64}$ of SO($6,6$) is not an irrep and it turns
out to be decomposed in terms of $\textbf{32}\,+\,\textbf{32'}$,
which represent left- and right-handed real spinors respectively
(Majorana-Weyl (M-W) spinors). This implies that, in an appropriate
basis\footnote{The one in which
$\Gamma^{13}=\Gamma^1\cdots\Gamma^{12}$ assumes the form
  diag($+\mathds{1}_{32},-\mathds{1}_{32}$)\,.}, one can introduce the
so-called 2-component formalism such that
\bea \chi^a=\left(\begin{array}{c} \chi^{\m}\\
\chi_{\dm}\end{array}\right)\,, \eea
where the index $\m$ lives in the $\textbf{32}$, whereas $\dm$ lives
in the $\textbf{32'}$. According to this decomposition, the Dirac
matrices will be decomposed into $32\times 32$ blocks as follows
\bea
\left[\Gamma^M\right]^{a}_{\phantom{a}b}=\left(\begin{array}{cc} 0 &
\left[\g^M\right]^{\m\dn}\\
\left[\bar{\g}^M\right]_{\dm\n} & 0
\end{array}\right) \ , \eea
and the charge conjugation matrix becomes
\bea \cC_{ab}=\left(\begin{array}{cc} \cC_{\m\n} & 0\\
0 & \cC^{\dm\dn}
\end{array}\right)\,. \eea
In terms of these $32\times 32$ gamma matrices, the relations
\eqref{Clifford} and \eqref{GammaT} can be, in the order, written as
follows
\bea
\left[{\g}^{(M}\right]^{\m\dot{\rho}}\,\left[\bar{\g}^{N)}\right]_{\dot{\rho}\n}=\eta^{MN}\,\delta^{\m}_{\phantom{a}\n}\hspace{5mm}\textrm{and}\hspace{5mm}
\left[\bar{\g}^M\right]_{\dm\n}=-
\left[{\g}^M\right]_{\n\dm}=-\,\cC_{\sigma \n}\left[{\g}^M\right]^{\sigma\dot{\rho}}\cC_{\dm \dot{\rho}}\,.\eea
All the antisymmetrised products of two gamma matrices can be
defined remaining either within the $\textbf{32}$ or the $\textbf{32'}$ in the following way
\bea
[\g^{MN}]^{\m}_{\phantom{a}\n}=[\g^{[M}]^{\m\dot{\rho}}\,[\bar{\g}^{N]}]_{\dot{\rho}\n}
\hspace{15mm} \textrm{and} \hspace{15mm}
[\g^{MN}]_{\dm}^{\phantom{a}\dn}=[\bar{\g}^{[M}]_{\dm\rho}\,[\g^{N]}]^{\rho\dn} \ .
\eea
As a consequence, one can generalise the above definitions to an
antisymmetrised product of an even number of gamma matrices, objects
which appear often in the present paper. Moreover, one can show that
in this type of representation
\bea \frac{1}{12!}\,\epsilon_{M_1 M_2\cdots M_{11}
M_{12}}\,[\g^{M_1}]^{\m_1\dot{\rho_1}}\,[\bar{\g}^{M_2}]_{\dot{\rho_1}\m_2}\cdots[\g^{M_{11}}]^{\m_5\dot{\rho_6}}\,[\bar{\g}^{M_{12}}]_{\dot{\rho_6}\m_6}=\delta^{\m_1}_{\phantom{ab}\m_6} 
\eea
and similarly for the dotted case but with the opposite sign. This means that the only independent irrep's of SO($6,6$) which can
be constructed by means of antisymmetrised products of gamma
matrices are all the $p$-forms up to degree six, the higher-degree
ones from 7 to 12 being just related to them by Hodge duality.
\emph{In particular the antisymmetrised product of six gamma
matrices turns out to be anti-selfdual} (ASD) \emph{when involving undotted indices and self-dual} (SD) \emph{when involving dotted indices}. After defining all
products of gamma matrices, one can make use of $\cC_{\m\n}$, $\cC^{\dm\dn}$ and their
inverse transpose in order to write those objects with two upper or
two lower indices. After doing this, one realises that
antisymmetrised products of \emph{two} and \emph{six} gamma matrices
are symmetric, whereas the ones with \emph{four}
are antisymmetric.

%
%

\bibliography{dgrN8}
\bibliographystyle{utphys}

\end{document}